\pdfoutput=1
\documentclass[aps,prl,twocolumn,superscriptaddress]{revtex4-2}

\usepackage{amsmath}
\usepackage{amssymb}
\usepackage{amsfonts}
\usepackage{amsthm}
\usepackage{amsbsy}
\usepackage{bm}
\usepackage{array}
\usepackage{mathtools}
\usepackage[usenames,dvipsnames]{xcolor}
\usepackage{dsdshorthand}
\usepackage{graphicx}
\usepackage[colorlinks=true,citecolor=blue,urlcolor=blue,linkcolor=blue,pdfstartview=FitH,bookmarksopen]{hyperref}
\usepackage{tikz}
\usepackage{comment}
\def\bpm{\begin{pmatrix}}
\def\epm{\end{pmatrix}}

\newcommand\vx{\mathbf{x}}
\newcommand\vk{\mathbf{k}}

\begin{document}

\title{Nonrelativistic Conformal Collider Physics of Multiparticle Point Production}

\author{Cyuan-Han Chang}
\affiliation{Leinweber Institute for Theoretical Physics, University of Chicago, Chicago, Illinois 60637, USA}

\author{Subham Dutta Chowdhury}
\affiliation{The Abdus Salam Institute for Theoretical Physics, Strada Costiera 11, 34151, Trieste, Italy}

\author{Ian Moult}
\affiliation{Department of Physics, Yale University, New Haven, Connecticut 06511, USA}

\author{Dam Thanh Son}
\affiliation{Leinweber Institute for Theoretical Physics, University of Chicago, Chicago, Illinois 60637, USA}

\date{August 2026}

\begin{abstract}
We define detector operators in the nonrelativistic conformal field theory describing fermions at unitarity.  We reduce the problem of computing the momentum distribution and correlation between final particles produced by a local source (``point-produced'') to the computation of correlation functions involving the detector operators. The general formalism is applied to the point production of three unitary fermions, where we find the momentum and angular distribution of final particles.  We discuss a nonrelativistic version of celestial holography, which maps the asymptotic out-state to a quantum wave function in the so-called ``oscillator frame.'' 
\end{abstract}

\maketitle

\emph{Introduction.}---Understanding systems
that consist solely of neutrons has been a long-standing problem in nuclear physics~\cite{Zeldovich:1960}.  Besides the intrinsic value, this problem is of direct relevance to the physics of neutron halo nuclei~\cite{Tanihata:1995yv}, nuclei near the neutron drip line~\cite{Nowacki:2021fjw}, and neutron star crusts~\cite{Chamel:2008ca}.  One approach to study such systems relies on  experiments that involve nuclear reactions in which several (typically from 2 to 6) neutrons are produced in the final state. Some examples of such reactions are radiative capture of $\pi^-$ by the triton ($\pi^-+{^3}\text{H}\to\gamma+3n$)~\cite{Miller:1980pn}, double charge exchange of $\pi^-$ on $^4$He ($\pi^-+{^4}\text{He}\to \pi^++4n$)~\cite{Ungar:1984ty}, and core knockout from $^8$He (e.g., $p+{^8}\text{He}\to p+{^4}\text{He}+4n$~\cite{Duer:2022ehf}). In these experiments, the kinematic region of special interest is where the relative energy~\footnote{The relative energy is the total kinetic energy of the neutrons in the frame moving with their common center of mass.} of the produced neutrons is small, typically a few MeV or less. At these energies the dynamics of the produced neutrons are crucially affected by their nonperturbative interactions, characterized, in particular, by an anomalously large scattering length $a\approx-19$~fm. One particularly intriguing question involves the possible existence of multi-neutron resonances.  The experimental situation with the tetraneutron is particularly interesting \cite{Kisamori:2016jie, Duer:2022ehf} (see also Ref.~\cite{Faestermann:2025our} for a recent review).

Recently, it has been shown~\cite{Hammer:2021zxb} that in the approximation when neutrons are regarded as fermions at unitarity (i.e., interacting through a zero-range interaction fine-tuned to infinite scattering length), the rate of production of $N$ neutrons grows as a function of the neutron relative energy $E$ as $E^{\Delta_N-5/2}$, where $\Delta_N$ is the scaling dimension of the lowest-dimensional operator creating $N$ particles in the nonrelativistic conformal field theory (NRCFT)~\cite{Nishida:2007pj} of unitary fermions.  This behavior can be considered a generalization of Wigner's threshold law~\cite{Wigner:1948zz} for unitary fermions and a nonrelativistic version of Georgi's ``unparticle physics''~\cite{Georgi:2007ek}. The correction to the power laws due to the finite scattering length was computed in Refs.~\cite{Chowdhury:2023oas, Beane:2025tum}, and the interpolation between the unitary-fermion and free-fermion behaviors was studied in Refs.~\cite{Backert:2026yjx,Higgins:2025dnz}.

Besides the total production rate, one can ask more detailed questions about the spectrum of the final particles and the correlation between them.  These observables can, in principle, be studied experimentally using modern detectors capable of registering multiple outgoing neutrons~\cite{Kobayashi:2013tha,R3B:2021lzy}. In the relativistic context, analogous quantities are studied in the modern approach to scatterings via energy correlators~\cite{Hofman:2008ar,Moult:2025nhu}. The energy correlators provide a bridge connecting quantum field theory quantities and collider observables.

The purpose of this Letter is to develop the theory of particle production based on the notion of \emph{detector operators} in NRCFT, mirroring the use of these operators in relativistic quantum field theory. In this approach, which we dub ``nonrelativistic conformal collider physics,'' the problem of characterizing the final state of a multiple-particle production process is reduced to the computation of one- or multi-point correlation functions of detector operators in a given initial state.  The detector operators form a new class of operators in NRCFTs: while local operators are functions of space and time coordinates, detector operators are functions of the \emph{velocity}. Physically, they can be identified with observables measured by an observer (detector) receding to infinity at a given velocity.  By expanding the notion of operators, quantities normally associated with the $S$-matrix become ordinary correlation functions in NRCFT.

We will also show that by using a special spacetime mapping---the map from the ``free-space frame'' to the ``oscillator frame''---one can map the scattering out-state to a quantum state, which arises as the result of a finite-time evolution in a harmonic trapping potential.  In this mapping, detector operators on the free-space side map to local operators on the oscillator side in such a way that the velocity variable in the detector operators becomes the coordinate variable in the local operators.  The spectrum of final particles over momentum becomes, upon the map, the density profile of a quantum state.  This correspondence between the asymptotic states and quantum wave functions can be thought of as a nonrelativistic version of ``celestial holography.''

\emph{Formulation of the problem.}---We first formulate the theoretical problem---an idealization of the physical scattering processes producing neutrons.  Consider a process, in which $N$ spin-$1/2$ particles with mass $m$ (which we will sometimes call ``neutrons'') are produced at a point with a given total energy $E$ and total momentum $\mathbf P$.  Thanks to Galilean invariance we can set $\mathbf P=\mathbf 0$ without losing generality. After being produced, the particles fly to infinity, their $N$-body quantum state evolving according to the Hamiltonian of unitary fermions (i.e., particles interacting through a zero-range potential fine-tuned to infinite scattering length).  The problem is to find the spectrum (distribution over momentum) and the correlations (two-particle, three-particle) between the momenta of the emitted neutrons.

There are two idealizations involved in formulating the above problem.  First, we assume that the particles are unitary fermions. For neutrons, this approximation is good when the relative energy between any pair of them is between $\hbar^2/m_n a^2\approx 0.1$~MeV and $\hbar^2/m_n r_\text{eff}^2\approx 5$~MeV (here $m_n$ is the neutron mass and $r_\text{eff}$ the effective range of neutron scatterings). Second, we assume that the neutrons are produced at a point, ignoring the size of the neutron and the finite extent of the initial nuclei from which the neutrons emerge. This is a good approximation when the wavelength of the final-state neutrons is large compared to the size of the source. 

The process of point production is characterized by (i) the primary operator $\mathcal O^\dagger$ that produces particles, and (ii) the total energy $E$ of the state.  We consider the following initial state,
\begin{equation}\label{eq:init-state}
  | \mathcal O, E \rangle = Z \! \int\!\mathrm dt\, \mathrm d \mathbf x\, \mathrm e^{- iEt}  \mathcal O^\dagger(t,\mathbf x) |0\rangle,
\end{equation}
normalized to unit norm by the prefactor $Z$.
In the standard scattering theory, one defines the form factor
\begin{equation}
   S(\vk_1, \vk_2, \ldots, \vk_N) = {}_\text{out}{}\< \vk_1, \vk_2, \ldots, \vk_N | \mathcal O, E\> .
\end{equation}
The single-particle spectrum of the outgoing particles, their two-particle correlation, etc., can be obtained by integrating $|S|^2$ over $N-1$, $N-2$, etc., of the $N$ momenta.  We now express these quantities in terms of the matrix elements of detector operators and their products.

\emph{Detector operators.}---In relativistic conformal collider physics~\cite{Hofman:2008ar} the detector operator $\mathcal E(\mathbf n)$ measures the total flux of energy flowing to infinity around the direction $\mathbf n$ ($\mathbf n^2=1$) per unit solid angle.  In NRCFT, 
detector operators are functions of not only the direction $\mathbf n$ but also of the velocity $v$, and the two variables can be combined into one vector velocity $\mathbf v=v\mathbf n$~\footnote{Cf.\ Ref.~\cite{Mateu:2012nk} where a velocity variable is introduced in detector operators of massive relativistic field theories.}. One imagines an experimental setup where a detector is located at a distance $R$ from the reaction point along the direction $\mathbf n$.  Ultrarelativistic final particles moving with the speed of light $c$ reach the detector at approximately the same time $R/c$, but nonrelativistic final particles reach the detector at all possible times. By measuring the mass flux (or energy flux) of particles that reach the detector at time $t=R/v$ we select those final particles moving along the direction $\mathbf n$ with velocity $v$~\footnote{For particles moving with velocity $v$ the energy flux and the mass flux differ by the factor of $v^2/2$, so in practice there is only one detector operator for the two fluxes.}.
We start with the mass density operator,
\begin{equation}\label{eq:rho-def}
  \rho(t,\mathbf x) = m\sum_{\alpha=\uparrow,\downarrow}\psi^\dagger_\alpha(t,\mathbf x) \psi_\alpha(t,\mathbf x),
\end{equation}
and define an operator $\mathcal M(\mathbf v)$ through a limiting procedure,
\begin{equation}\label{eq:Mv-def}
  \mathcal M (\mathbf v) = \lim_{t\to+\infty} t^3 \rho (t, \mathbf vt).
\end{equation}
The factor $t^3$ on the right-hand side is introduced so that the integral of
$\mathcal M(\mathbf v)$ over $\mathbf v$ is the total mass of the system,
\begin{equation}\label{eq:total-M}
   M = \int\! \mathcal M (\mathbf v)\, \mathrm d\mathbf v
   = \int\! \mathcal M (\mathbf v)\, v^2 \mathrm dv\, \mathrm d\mathbf n
   .
\end{equation}
We now argue that one can interpret $\mathcal M(\mathbf v)$ as a detector operator, and $\mathcal M(\mathbf v)\, v^2\mathrm dv\, \mathrm d\mathbf n$ is the mass flux that comes from particles with velocity in the range $(v,v+\mathrm dv)$ that pass through a detector located at direction $\mathbf n$ and covering the solid angle $\mathrm d\mathbf n$. The right-hand side of Eq.~(\ref{eq:total-M}) is then simply the total mass that passes through all detectors covering the full $4\pi$ solid angle by particles of all possible velocities.

To demonstrate that identification, we introduce the mass flux operator $\mathbf j(t,\mathbf x)=-\frac i2 \sum_{\alpha=\uparrow,\downarrow}(\psi^\dagger_\alpha \bm{\nabla}\psi_\alpha-\bm{\nabla}\psi^\dagger_\alpha\psi_\alpha)$. Assume we have a detector located at a large distance $R$ along the direction $\mathbf n$ from the interaction point, covering a solid angle $\mathrm{d}\mathbf n$.  The mass passing through that solid angle during the time interval $(t, t+\mathrm dt)$ is $\mathrm d M= \mathbf n\cdot \mathbf j(t, R\mathbf n)\, R^2\, \mathrm dt\, \mathrm d\mathbf n$.  Changing variable from $t$ to $v=R/t$, this becomes $\mathrm dM=\mathbf n\cdot \mathbf j(R/v, R\mathbf n)\, R^3\, v^{-2}\, \mathrm dv\, \mathrm d\mathbf n$.
In the limit $t,R\to\infty$, $R/t=\text{fixed}$, we select out particles with velocity $v\mathbf n$ where $v=R/t$, and we can replace,
\begin{equation}\label{eq:j=rhov}
   \mathbf n \cdot \mathbf j\Bigl(\frac Rv, R\mathbf n\Bigr) 
   \to v \rho\Bigl(\frac Rv, R\mathbf n\Bigr)  \,.
\end{equation}
Now using Eq.~(\ref{eq:Mv-def}) we find $\mathrm dM=\rho(t,\mathbf vt)v^2t^3 \mathrm dv\, \mathrm d\mathbf n$.

Equation~(\ref{eq:j=rhov}) allows us to write an alternative definition for the detector operator $\mathcal M(\mathbf v)$:
\begin{equation}
  \mathcal M(\mathbf v) = \lim_{t\to\infty}
  \frac{t^3}{v^2}\mathbf v\cdot \mathbf j(t,\mathbf vt).
\end{equation}

In addition to Eq.~(\ref{eq:total-M}), physical consideration implies that the total energy $E$ of the system can also be expressed as an integral,
\begin{equation}\label{eq:E-conserv}
  E = \int\! \mathcal M(\mathbf v) \frac{v^2}2 \, \mathrm d\mathbf v .
\end{equation}

If the theory has only one particle type, then the full momentum-space probability distribution of the final state is contained in the matrix element of $\mathcal M$ or its products.  For example, the momentum distribution function of the final particles is
\begin{equation}
  \int\!\mathrm d\vk_2\ldots \mathrm d\vk_N \, |S(\vk, \vk_2,\ldots, \vk_N)|^2  \sim \langle \mathcal M(\mathbf v)\rangle ,
\end{equation}
where $\vk=m\mathbf v$ and the average on the right-hand side is taken over the state $|\mathcal O, E\rangle$ given in Eq.~(\ref{eq:init-state}).  Similarly, the correlation between two final particles in momentum space is given by the connected two-point correlator of the detector operator $\mathcal M(\mathbf v)$.

\emph{Nonrelativistic ``celestial holography.''}---There is a free space--harmonic trap correspondence, which maps the time evolution of the system from the initial state at $t=0$ to the out-state at $t=\infty$ to a \emph{finite-time} evolution, in an isotropic harmonic potential of frequency $\Omega$, from the same initial state at ``oscillator time'' $\tau=0$ to a final state at $\tau=\pi/(2\Omega)$. The coordinates of the ``oscillator frame'' $\tau$ and $\mathbf y$ are related to the free-space coordinates through~\cite{Niederer:1973tz,Werner:2006zzb,Goldberger:2014hca},
\begin{equation}\label{eq:oscillator_transform}
  \Omega t = \tan\Omega\tau, \qquad \mathbf x = \frac{\mathbf y}{\cos\Omega\tau}\,,
\end{equation}
where $\Omega$ is an arbitrary parameter.  Note that the whole half-line $0<t<\infty$ maps to a finite interval $0<\tau<\pi/(2\Omega)$.
The exact correspondence between free-space and oscillator time evolutions works as follows.  Given a solution $\Psi(t,\mathbf x_a)$ to the many-body Schr\"odinger equation, 
\begin{equation}
   i \frac{\partial}{\partial t}\Psi (t, \mathbf x_a) = H \Psi(t,\mathbf x_a),
\end{equation}
with the interacting Hamiltonian,
\begin{equation}
  H = -\frac1{2m} \sum_a \frac{\partial^2}{\partial\mathbf x_a^2}
  + \sum_{a<b} V(\mathbf x_a - \mathbf x_b),
\end{equation}
one can construct a function of the oscillator-frame coordinates, $\tilde\Psi(\tau, \mathbf y_a)$, through,
\begin{equation}\label{eq:map-wf}
  \Psi(t,\mathbf x_a) = \cos^{\frac32\!N}\Omega\tau \exp\biggl( \frac i2 m\Omega\tan\Omega\tau\sum_a y_a^2\biggr)\tilde\Psi(\tau,\mathbf y_a).
\end{equation}
Defined this way, $\tilde\Psi$ satisfies the Schr\"odinger equation of a system of particles in an isotropic harmonic trapping potential, interacting with each other through a time-dependent interparticle potential: 
\begin{equation}
      i \frac{\partial}{\partial \tau}\tilde\Psi (\tau, \mathbf y_a) =  H_\text{osc} \tilde\Psi(\tau,\mathbf y_a),
\end{equation}
where, 
\begin{equation}
   H_\text{osc} = \sum_a \left(-\frac1{2m} \frac{\partial^2}{\partial\mathbf y_a^2} + \frac{m\Omega^2}2 y_a^2\right) + \sum_{a<b}\tilde V(\tau; \mathbf y_a -\mathbf y_b),
\end{equation}
and the interparticle potential $\tilde V$ is the rescaled version of the original potential $V$ with a time-dependent scale factor $\lambda=1/\cos\Omega\tau$,
\begin{equation}
  \tilde V(\tau, \mathbf X) = \frac1{\cos^2\Omega\tau} V\Bigl(\frac{\mathbf X}{\cos\Omega\tau} \Bigr).  
\end{equation}
The rescaling is done in such a way that if $V(\mathbf X)$ supports a bound state at threshold, then $\tilde V$ also does, for any $\lambda>0$.  Taking the range of $V$ to zero, one then maps the evolution of a system of unitary fermions in free space to the evolution of the same system in a harmonic trap.

The oscillator frame is useful for our purposes because, using Eqs.\ \eqref{eq:rho-def}, \eqref{eq:Mv-def}, \eqref{eq:oscillator_transform}, and \eqref{eq:map-wf}, the detector operator in free space becomes the density operator at the ``end of oscillator-frame time'':
\begin{equation}
  \mathcal M(\mathbf v) = \frac1{\Omega^3}\tilde\rho\left(\frac\pi{2\Omega}, \frac{\mathbf v}\Omega \right).  
\end{equation}
where $\tilde\rho(\mathbf y) =m\sum_a\delta(\mathbf y-\mathbf y_a)$. Thus the correlators of the final particles in point production, which ultimately are properties of a form factor, are mapped to correlators between local operators in a nonrelativistic conformal field theory in the presence of a harmonic potential, in such a way that velocity in free space is mapped to  
coordinates in the harmonic trap~\footnote{A similar correspondence is observed when a unitary Fermi gas is released from an isotropic harmonic trap: the momentum distribution of final particles is homothetic to the density distribution of the initial state~\cite{2012LNP...836..127C}.}.
This correspondence is similar in spirit to the goal of the ``celestial holography'' program~\cite{Pasterski:2021raf}. Note, however, that our correspondence is not a ``holography'' in the true sense---the NRCFT in the oscillator frame lives in the same number of dimensions as the original theory.  On the other hand, the physical observables live on one particular time slice.

We summarize the correspondence between the two frames in Table~\ref{table:celestial}.
\begin{table}[ht]
\begin{center}
\begin{tabular}{|c||c|}
\hline
Free-space frame & Oscillator frame\\
\hline
asymptotic out-state & state at surface $\tau=\pi/2\Omega$\\
\hline
velocity $\mathbf v$ & position $\mathbf y=\mathbf v/\Omega$\\
\hline
 $\,$detector operator $\mathcal M(\mathbf v)\,$ & $\,$density operator $\tilde\rho(\pi/2\Omega, \mathbf y)$\\ 
\hline
\end{tabular}
\end{center}
\caption{Nonrelativistic ``celestial holography.''}
\label{table:celestial}
\end{table}

\emph{Point production of three unitary fermions.}---To illustrate the use of detector operators in NRCFT, we compute the spectrum of final particles when three unitary fermions are produced at a point.  The problem reduces to the computation of the three-point function,
\begin{equation}
   \int\!\mathrm dt\, \mathrm d\mathbf x\, \mathrm e^{ iEt} \langle 0 | \mathcal O(t,\mathbf x) \mathcal M(\mathbf v) \mathcal O^\dagger(0,\mathbf 0) | 0 \rangle .
\end{equation}

The lowest-dimensional charge-3 operator is the $l=1$ operator with dimension $\Delta=4.27\ldots$~\cite{Nishida:2007pj}.  Assuming that $\mathcal O$ is polarized along the $z$-direction, the particle spectrum has the following dependence on the direction of $\mathbf v$,
\begin{equation}
  \langle \mathcal M(\mathbf v) \rangle = \frac{m}{v_0^3} \rho_{M}\Bigl(\frac{v}{v_0}\Bigr) \left[ 1 +  \alpha_2\Bigl(\frac v{v_0}\Bigr) \left(\cos^2\theta-\frac13\right) \right],   
\end{equation}
where $v_0=\sqrt{E/m}$, $\theta$ is the angle between $\mathbf v$ and the $z$-axis, and $\rho_{M}(\tilde v)$ and $\alpha_2(\tilde v)$ are dimensionless functions of $\tilde v=v/v_0$. The maximal kinematically allowed value of $\tilde v$ is $2/\sqrt3$.  The function $\rho_M(\tilde v)$ satisfies the sum rules that follow from Eqs.~(\ref{eq:total-M}) and (\ref{eq:E-conserv}): 
\begin{equation}
  \int\limits_0^\infty\! d\tilde v\, 4\pi \tilde v^2\rho_M(\tilde v) = 3, \quad
  \int\limits_0^\infty\! d\tilde v\, 2\pi \tilde v^4\rho_M(\tilde v) = 1 ,
\end{equation}
while positivity of $\<\mathcal M(\mathbf v)\>$ requires that $\alpha_2(\tilde v)$ satisfies the Hofman--Maldacena bound $-\frac32\leq\alpha_2\leq3$~\cite{Hofman:2008ar}.

\begin{figure}[ht]
  \includegraphics[width=0.95\columnwidth]{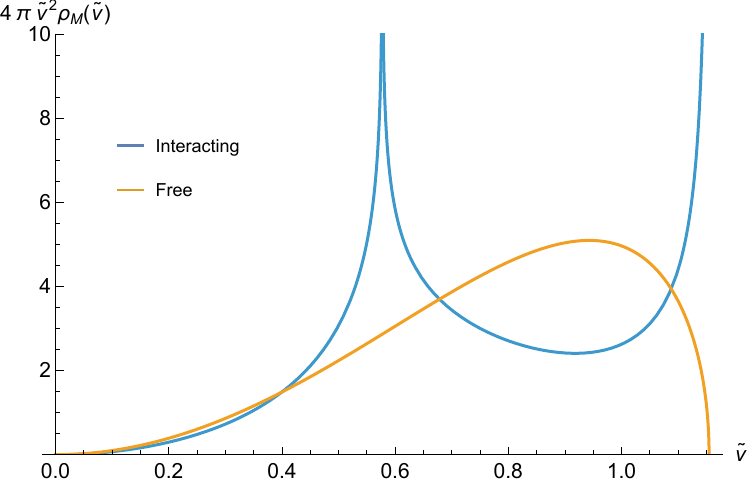}
  \caption{Plot of $4\pi \tilde v^2\rho_M(\tilde v)$ (the distribution of final particles over velocity) for unitary fermions (blue line) and free fermions (orange line). The areas below the curves are 3 (the number of final particles).  For unitary fermions the function becomes singular at $\tilde v=1/\sqrt3$ and $2/\sqrt3$ (see text).}
  \label{fig:rhospin1} 
\end{figure}

\begin{figure}[ht]
  \includegraphics[width=0.95\columnwidth]{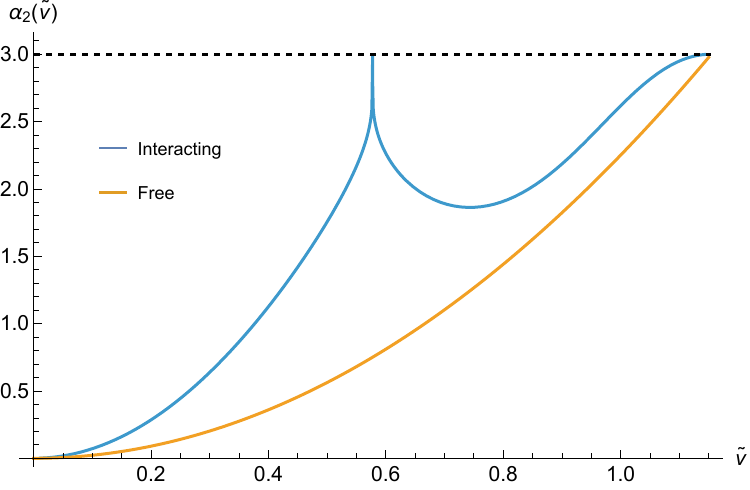}
  \caption{Plot of $\alpha_2(\tilde v)$ for unitary fermions (blue line) and free fermions (orange line). The dashed line is the upper Hofman--Maldacena bound, saturated at $\tilde v=1/\sqrt3$ and $2/\sqrt3$.}
  \label{fig:alpha2spin1}
\end{figure}

The calculation is performed by inserting and summing over a complete set of 3-body scattering states between the detector operator and the operator that acts on the vacuum to produce the particles~\cite{SM}.  The results for the functions $\rho_M(\tilde v)$ and  $\alpha_2(\tilde v)$ are plotted in Figs.~\ref{fig:rhospin1} and \ref{fig:alpha2spin1}.  For comparison, we also plot the same functions for free spin-$1/2$ fermions, where they can be computed analytically,
\begin{equation}\label{eq:rhoMv2-free}
  \rho_{M}^\text{free}(\tilde v) = \frac{9\sqrt3}{4\pi^2}\sqrt{4-3\tilde v^2}\,, \quad
  \alpha_2^\text{free}(\tilde v) = \frac94\tilde v^2.
\end{equation}
The singularities in the function $\rho_{M}(v)$ are due to the configuration where two of the outgoing particles with opposite spins have the same velocities. In this configuration one particle has velocity close to $2v_0/\sqrt{3}$ and two other particles have their velocities close to $v_0/\sqrt3$.  The resonance interaction within the pair of particles with almost the same momentum leads to the enhancement of the distribution of particles near these values of the velocity.  It can be shown that the singularities of $\rho_M(\tilde v)$ are of the form $(2/\sqrt 3-\tilde v)^{-1/2}$ and $\ln|1/\sqrt{3}-\tilde v|$ near the two singularities.  The singularity near the end point $2v_0/\sqrt{3}$ is the Watson-Migdal singularity~\cite{Watson:1952ji, Migdal:1955}.  The logarithmic singularity at $v=v_0/\sqrt{3}$ is more subtle: it appears when one fixes one particle's momentum to be near $v_0/\sqrt{3}$ and integrates over the momenta of two other particles, picking up the singular behavior from the region of integration where one of the remaining two particles with opposite spin also has its momentum near $v_0/\sqrt3$ and the direction of $\mathbf{v}$.

An analytic approximation to $\rho_M(\tilde v)$ is given in the End Matter.

\emph{Point production of a large number of unitary fermions.}---For the production of four and more particles, we do not have the exact few-body wave functions.  However, when the number of produced particles is large, analytic calculation is again possible as the many-body system can now be described by the superfluid effective field theory.  The path integral that determines expectation value of $\mathcal M(\mathbf v)$ is saturated by the saddle-point found in Ref.~\cite{Beane:2024kld}~\footnote{For the relativistic case, see Ref.~\cite{Cuomo:2025pjp}.}. 
The result reads,
\begin{equation}\label{eq:rhoM-largeN}
   \langle \mathcal M(\mathbf v) \rangle = \frac{27(Nm)^4}{512\pi^2 E^3}\left(\frac{16E}{3Nm} -v^2\right)^{3/2},
\end{equation}
for $v<v_\text{max}=\sqrt{16E/(3Nm)}$ and is 0 when $v>v_\text{max}$. This coincides with the spectrum for free fermions, which was found by a completely different argument in Ref.~\cite{Son:2021kkx}. Note that the spectrum has a cutoff at $v=v_\text{max}\sim \sqrt{E/(Nm)}$ while the maximally allowed velocity of an outgoing particle should be $\sqrt{2E/m}$.  The sharp cutoff at $v=v_\text{max}$ is the result of taking the large $N$ limit; one expects the real spectrum to have a tail beyond $v=v_\text{max}$ when one goes beyond the leading order in $1/N$.

\emph{Conclusion.}---In this Letter we have introduced the notion of detector operators in NRCFT and used it to understand multiparticle point production. As an illustration, we have computed the single-particle spectrum in the process of production of three particles in the theory of unitary fermions.  It would be interesting to compare this result with experimental data~\cite{Miki:2024}.

One topic that we have left for future work is inclusion of a finite scattering length.  The problem is still a well-defined one, but instead of an NRCFT we have a theory undergoing a renormalization-group flow from two fixed points. It was found that the scattering length is numerically quite important for the total rate of point production of neutrons~\cite{Chowdhury:2023oas,Beane:2025tum,Higgins:2025dnz}.  We expect that it will smooth out the singularities in the spectrum of the final particles. It would be interesting to study whether the idea of detector matching \cite{Chang:2025zib} can be useful for this problem. It would also be of interest, given the recent experiments~\cite{Faestermann:2025our,Nasr:2025lmi} to develop techniques to tackle point production of four and more neutrons.

Another important problem is to understand how to characterize the finite size of the initial state, as it is the case with neutrons emerging from collision of nuclei.  It has been argued~\cite{Lazauskas:2022mvq} that the peak observed in the $4n$ spectrum in the core knockout experiment with $^8\text{He}$ beam~\cite{Duer:2022ehf} cannot be explained without correlation between the neutrons in the initial $^8\text{He}$ nucleus. One may speculate that the finite size of the initial state requires a formalism that includes all primary operators with a given mass, not just the one with the lowest dimension.

Finally, we note that standard measurements, which determine the probability distribution of final particles over the momentum, are sensitive only to the absolute value squared of the oscillator-frame wave function, but not its phase. This raises an interesting question: can one measure the phase of the oscillator-frame wave function?  Are quantum characteristics of that wave function (e.g.,  bipartite entanglement entropy), observable?

\acknowledgments 

The authors thank Clay Córdova, Gabriel Cuomo, Hans-Werner Hammer, Justin Kulp, Jonathan Sorce, Jesse Thaler and F\'elix Werner for discussions. The authors thank organizers of ``Bootstrap 2025'' where this work was initiated. The work of D.T.S. is supported, in part, by the U.S.\ DOE
Grant No. DE-FG02-13ER41958.  D.T.S. thanks the Institut de Hautes \'Etudes Scientifiques, where part of this work was completed, for hospitality. IM is supported by the DOE Early Career Award DE-SC0025581, and the Sloan Foundation.
S.D.C.\ is supported by ``Exotic High Energy Phenomenology'' (X-HEP), a project funded by the European Union -- Grant Agreement n.~101039756 (PI: J.~Elias~Mir\'o). Views and opinions expressed are however those of the author(s) only and do not necessarily reflect those of the European Union or the ERC Executive Agency (ERCEA). Neither the European Union nor the granting authority can be held responsible for them.

\bibliography{refs}

\section*{End Matter: Approximate analytic expression for $\rho_M(\tilde v)$
}
\label{app:approx}

\setcounter{equation}{0}
\renewcommand{\theequation}{A\arabic{equation}}

To facilitate potential comparison with  experiment, we give here an approximate analytic expression for $\rho_M(\tilde v)$. The final particle spectrum over velocity $\rho_M(\tilde v)$ can be approximated,  
with an accuracy uniformly better than 
1\%, by the following function~\footnote{The fit was suggested by Claude AI (Opus 5) and verified by hand.}:
\begin{multline}\label{eq:rhoM-approx}
 \rho_M(\tilde v) \approx w(\tilde v) \biggl[
 \frac{A_1}{w^2(\tilde v)} + \frac{A_2}{\tilde v w(\tilde v)} \ln \left| \frac{w(\tilde v) + \frac32\tilde v}{w(\tilde v) - \frac32\tilde v} \right|\\ + c_0 
 + c_1 \tilde v^2 + c_2 \tilde v^4 
 \biggr],
\end{multline}
with,
\begin{equation}
  w(\tilde v) \equiv \sqrt{1-\frac34 \tilde v^2}  \,,
\end{equation}
$A_1$ and $A_2$ fixed to reproduce the magnitude of the singular part of $\rho_M(\tilde v)$ at $\tilde v=2/\sqrt3$ and $1/\sqrt3$,
\begin{equation}
  A_1 = 0.086958,\quad A_2 =0.231708,
\end{equation}
and $c_{0,1,2}$ being fit parameters,
\begin{equation}
  c_0 = -0.22713, \\ \quad c_1 = -0.08817, \quad c_2 = 0.06295
.  
\end{equation}

\clearpage
\onecolumngrid

\begin{center}
        \textbf{\large --- Supplemental Material ---\\ $~$ \\
        Nonrelativistic Conformal Collider Physics of Multiparticle Point Production}\\
        \medskip
        \text{Cyuan-Han Chang, Subham Dutta Chowdhury, Ian Moult, and Dam Thanh Son}
\end{center}
\setcounter{equation}{0}
\setcounter{figure}{0}
\setcounter{table}{0}
\setcounter{page}{1}
\makeatletter
\renewcommand{\thesection}{S\arabic{section}}
\renewcommand{\theequation}{S\arabic{equation}}
\renewcommand{\thefigure}{S\arabic{figure}}
\renewcommand{\bibnumfmt}[1]{[S#1]}

\section{Calculations of $\rho_M(v)$ and $\alpha_2(v)$}

In this Supplemental Material we sketch out the computation of the spectrum of final particles produced by acting the lowest-dimensional charge-3 operator on the vacuum.  We will set $m=1$.

In the theory of unitary fermions, for each angular momentum channel with quantum numbers $(lm)$ there is a tower of 
charge-3 operators~\cite{Chowdhury:2023oas}.  Each operator in the tower is labeled by three indices $l$, $m$, $s$: $\mathcal O^{l,m,s}$, where $s$ is related to the dimension of the operator by $\Delta=\frac52+s$.  The allowed values for $s$ depend on $l$; for $l=0$ and $l=1$ the equations that determine the allowed $s$ are written below in Eqs.~(\ref{eq:phil01}) and (\ref{eq:eq-phi}).

Acting $O^{l,m,s\dagger}$ on the vacuum one obtains states of the form~\cite{Chowdhury:2023oas},
\begin{equation}\label{eq:3body-wf}
\Bigl|\Psi^{l,m,s}_{\mathbf P_{\textrm{cm}},\kappa}\Bigr\> = \frac{1}{\sqrt{2}}\int\! \mathrm d\mathbf x_1\, \mathrm d\mathbf x_2\, \mathrm d\mathbf x_3\, \Psi^{l,m,s}_{\mathbf P_{\textrm{cm}},\kappa}(\mathbf x_i)\psi^{\dagger}_{\uparrow}(\mathbf x_1)\psi^{\dagger}_{\downarrow}(\mathbf x_2)\psi^{\dagger}_{\uparrow}(\mathbf x_3)|0\>,
\end{equation}
with the wave function, 
\begin{equation}
\Psi^{l,m,s}_{\mathbf P_{\text{cm}},\kappa}(\mathbf x_i) = \left(\frac{2}{\sqrt{3}}\right)^{3/2}N^l_{\kappa}\, \mathrm e^{\mathrm i \mathbf P_{\text{cm}}\cdot\mathbf R_{\text{cm}}} \frac{J_s(\sqrt{2}\kappa R)}{R^2}\Phi^{l,m}_s(\Omega).
\end{equation}
Here one defines the Jacobi coordinates,
\begin{equation}\label{jacobi_app_1}
\mathbf R_{\text{cm}} = \frac{\mathbf x_1 + \mathbf x_2 + \mathbf x_3}{3}, \quad \mathbf r = \mathbf x_2 - \mathbf x_1, \quad \bm{\rho} =\frac{2}{\sqrt{3}}\left(\mathbf x_3 - \frac{\mathbf x_1 + \mathbf x_2}{2}\right),
\end{equation}
and through them the hyperradius $R$ and hyperangles $\Omega$,
\begin{equation}\label{jacobi_app_2}
R = \sqrt{\frac{r^2 + \rho^2}{2}},\quad \Omega=(\alpha,\hat{\mathbf r},\hat{\bm{\rho}}),\quad \alpha=\arctan\frac{r}{\rho}.
\end{equation}

The hyperangular part of the wavefunction is,
\begin{equation}
\Phi^{l,m}_s(\Omega) = (1-P_{13})\frac{\f_s^l(\a)}{\sin(2\a)}Y_l^m(\hat {\bm{\rho}}),
\end{equation}
where $Y_l^m(\hat {\bm{\rho}})$ is the spherical harmonics, $P_{13}$ is the operator that exchanges $\mathbf x_1$ and $\mathbf x_3$, and $\phi_s^l(\alpha)$ for $l=0$ and $l=1$ are,
\begin{equation}\label{eq:phil01}
  \f_s^{l=0}(\a)=\sin\Bigl[s\Bigl(\frac{\pi}{2}-\a\Bigr)\Bigr],\quad \f_s^{l=1}(\a) = -s\cos\Bigl[s\Bigl(\frac{\pi}{2}-\a\Bigr)\Bigr]+ \tan \a \sin\Bigl[s\Bigl(\frac{\pi}{2}-\a\Bigr)\Bigr].
\end{equation}
The allowed values of $s$ are determined by solving the equations
\begin{equation}\label{eq:eq-phi}
  \phi_s^{l\prime} (0) - (-1)^l \frac{4}{\sqrt3}\phi^l_s\Bigl(\frac\pi 3\Bigr) = 0.
\end{equation}
The smallest nontrivial solution for $l=0$ is $s=2.16622$ and for $l=1$ is $s=1.77272$.

We normalize $\Psi^{l,m,s}_{\mathbf P_\text{cm}, \kappa}$ so that, 
\begin{multline}
   \Bigl\< \Psi^{l',m',s'}_{\mathbf P_\text{cm}',\kappa'}\Big| \Psi^{l,m,s}_{\mathbf P_\text{cm},\kappa} \Bigr\> =
    \int\!\mathrm d\mathbf x_1\, \mathrm d\mathbf x_2\, \mathrm d\mathbf x_3\, \Psi^{l',m',s'*}_{\mathbf P_\text{cm}',\kappa'}(\vx_1, \vx_2, \vx_3) \Psi^{l,m,s}_{\mathbf P_\text{cm},\kappa}(\vx_1, \vx_2, \vx_3)\\
    = (2\pi)^3\delta(\mathbf P_\text{cm}-\mathbf P_\text{cm}')\, \pi \delta \big(\sqrt2(\kappa-\kappa')\big)\, \delta_{ll'}\delta_{mm'}\delta_{ss'} .
\end{multline}
Then the normalization condition requires,
\begin{equation}
|N_\kappa^l|^2 = \frac{\sqrt{2}\pi\kappa}{f_s^l}\,,
\end{equation}
where, 
\begin{equation}\label{eq:fs_def}
f_s^l\equiv \int\! \mathrm d\bar \Omega\, |\Phi^{l,m}_s(\Omega)|^2,
\end{equation}
and $\mathrm d\bar\Omega\equiv 2\sin^22\alpha\,\mathrm d\alpha\, \mathrm d\hat{\bm \rho}\,\mathrm d\hat{\mathbf r}$.  

The state~(\ref{eq:3body-wf}) is an eigenstate of the Hamiltonian,
\begin{equation}
  H \Big|\Psi^{l,m,s}_{\mathbf P_\text{cm},\kappa}\Bigr\> = \left( \frac{P_\text{cm}^2}6 + \kappa^2\right)  \Big|\Psi^{l,m,s}_{\mathbf P_\text{cm},\kappa}\Bigr\> .
\end{equation}
In the Schr\"odinger picture the wavefunction at time $t$ is the wavefunction at $t=0$ times $\exp[-i(\frac16P_\text{cm}^2+\kappa^2)t]$.  In the detector limit (i.e., $t\to+\infty$, $\vx\to\infty$, $\vx/t=\text{const}$) the wavefunction can be simplified by using the following formulas, which are valid in the sense of distributions,
\begin{subequations}\label{eq:detector-lim}
\begin{align}
  \exp \left( - i\frac{P_\text{cm}^2}6t +  i \mathbf P_\text{cm}\cdot \mathbf R_\text{cm} \right) & \approx \left( \mathrm e^{-i \pi/4} \sqrt{\frac{6\pi}t}\right)^3 \exp\left( \frac{3i}2 \frac{R_\text{cm}^2}t \right) \delta^3 \left( \mathbf P_\text{cm} - \frac{3 \mathbf R_\text{cm}}t\right), \\
  \mathrm e^{-i\kappa^2t} J_s(\sqrt2 \kappa R) & \approx \frac{\mathrm e^{-\frac{i \pi}{2} (s+1)}  }{\sqrt{2} R} \exp\left(\frac{i R^2}{2 t}\right) \delta \left( \kappa - \frac{R}{\sqrt{2} t}\right).
\end{align}
\end{subequations}

The expectation value of the density operator can be written as,
\begin{equation}
   \<\Psi^{l',m',s'}_{\mathbf P_\text{cm}',\kappa'}|  \rho(t,\mathbf x) |\Psi^{l,m,s}_{\mathbf P_\text{cm},\kappa}\> = \int\!\mathrm d\mathbf x_1\, \mathrm d\mathbf x_2\, \bigl[ 2 \Psi^{\prime *}(\vx_1,\vx_2,\vx)\Psi(\vx_1, \vx_2, \vx) + \Psi^{\prime *}(\vx_1,\vx,\vx_2)\Psi(\vx_1, \vx, \vx_2) \bigr],
\end{equation}
where, on the right-hand side, $\Psi\equiv \Psi^{l,m,s}_{\mathbf P_\text{cm},\kappa}$ and $\Psi'\equiv \Psi^{l',m',s'}_{\mathbf P_\text{cm}',\kappa'}$, and we have used the antisymmetry of the wave functions $\Psi(\vx_1,\vx_2,\vx_3)$ and $\Psi'(\vx_1,\vx_2,\vx_3)$ under the exchange of $\vx_1$ and $\vx_3$. Taking the detector limit using Eqs.~(\ref{eq:detector-lim}), one can find the spectrum of final particles which one gets by acting $\mathcal O^{l,m,s,\dagger}$ on the vacuum. 

For $l=1$ one finds,
\begin{align}
  \rho_M(\mathbf v) &= \frac{\sqrt{3}}{f^{1}_s}\sqrt{1-\frac{3v^2}{4}}  [I_0(\mathbf{v})+2I_1(\mathbf{v})]\,, \\
  \alpha_2(\mathbf{v}) &=\frac{3[I_0(\mathbf{v})-I_1(\mathbf{v})]}{I_0(\mathbf{v})+2I_1(\mathbf{v})},
\end{align}
where,
\begin{equation}\label{eq:Imv}
    I_m(\mathbf{v})=\left.\int\! \mathrm d^2\hat{\mathbf q}\left(2\left|\Phi^{1,m}_s(\Omega_{\mathbf{y},\mathbf{z},\mathbf v})\right|^2+\left|\Phi^{1,m}_s(\Omega_{\mathbf{y},\mathbf v,\mathbf{z}})\right|^2\right)\right|_{\mathbf{y} =\hat{\mathbf{q}}\sqrt{1 - \frac{3}{4}v^2} -\frac12{\mathbf v},~ \mathbf{z} = -\hat{\mathbf{q}}\sqrt{1 - \frac{3}{4}v^2}  -\frac12{\mathbf v}} \,,
\end{equation}
where $\Omega_{\mathbf{a},\mathbf b,\mathbf{c}}$ are the hyperangles obtained by setting $\mathbf{x_1}=\mathbf{a}$, $\mathbf{x_2}=\mathbf{b}$ and $\mathbf{x_3}=\mathbf{c}$ in Eqs.\ \eqref{jacobi_app_1} and \eqref{jacobi_app_2}. Evaluating the integrals in Eq.~(\ref{eq:Imv}) numerically for the lowest value of $s$, one obtains results plotted in Figs.~\ref{fig:rhospin1} and \ref{fig:alpha2spin1}.  Repeating the calculation for the free field theory one obtains Eq.~(\ref{eq:rhoMv2-free}).

\end{document}